\documentclass[11pt]{article}
\usepackage{hyperref}
\usepackage{mathpazo}
\usepackage{amssymb,amsmath,amsthm}
\usepackage{epsfig}
\usepackage{tcolorbox}
\usepackage{enumitem}
\usepackage[T1]{fontenc}

\tcbset{ 
  halign=justify, 
  center, 
  colback=gray!10, 
  colframe=black, 
  boxrule=0.5pt 
}

\newtheorem{theorem}{Theorem}[section]
\newtheorem{proposition}[theorem]{Proposition}
\newtheorem{definition}[theorem]{Definition}

\newtheorem{lemma}[theorem]{Lemma}

\newtheorem{corollary}[theorem]{Corollary}

\def\lsoft{{l\kern-0.035cm\char39\kern-0.03truecm}}

\newcommand{\qCol}{\ensuremath{q}\textnormal{\textsc{-Coloring}}}
\newcommand{\ListCol}{\textnormal{\textsc{-List Coloring}}}
\newcommand{\qListCol}{\ensuremath{q}\textnormal{\textsc{-List Coloring}}}
\newcommand{\Empty}{\textsc{Empty}}
\newcommand{\coNPpoly}{\mathsf{coNP/poly}}

\newcommand{\qedsymb}{\hfill{\rule{2mm}{2mm}}}
\renewenvironment{proof}[1][]{\begin{trivlist}
\item[\hspace{\labelsep}{\bf\noindent Proof#1:\/}] }{\qedsymb\end{trivlist}}

\def\calG{{\cal G}}
\def\calF{{\cal F}}

\def\calP{{\cal P}}

\def\R{\mathbb{R}}

\def\N{\mathbb{N}}

\newcommand\Prob[2]{{\Pr_{#1}\left[ {#2} \right]}}

\newcommand{\NP}{\mathsf{NP}}

\newcommand{\YES}{\mathsf{YES}}
\newcommand{\NO}{\mathsf{NO}}
\newcommand{\Split}{\textsc{Split}}
\newcommand{\USplit}{\cup \textsc{Split}}

\newcommand{\eps}{\epsilon}
\renewcommand{\epsilon}{\varepsilon}

\begin{document}

\title{{\bf Setwise Distinguishable Permutations}}

\author{
Ishay Haviv\thanks{The Academic College of Tel Aviv-Yaffo, Tel Aviv, Israel.}
}

\date{}

\maketitle
\thispagestyle {empty}

\begin{abstract}
A family of permutations of $[n]$ is called {\em setwise distinguishable} if for every permutation in the family there exists a subset of $[n]$ whose image under this permutation differs from its image under any other permutation in the family. We prove that there exists a setwise distinguishable family of $2^{(2-o(1)) \cdot n}$ permutations of $[n]$. The result is optimal up to the $o(1)$ term in the exponent and is achieved through an explicit construction. As an application, we obtain nearly tight conditional lower bounds on the kernelization complexity of graph coloring problems parameterized by the vertex-deletion distance to split graphs. This improves a result of Jansen and Kratsch (Inf. Comput., 2013).
\end{abstract}


\section{Introduction}

Extremal combinatorics is the area that investigates the maximum and minimum possible sizes of families of discrete objects subject to prescribed constraints (see, e.g.,~\cite{Jukna11}). Beyond their intrinsic interest, such problems naturally arise in diverse fields, including computer science, information theory, number theory, and geometry. A central branch of this area is extremal set theory, which focuses on collections of sets and of objects that involve sets in various ways. One cornerstone result, due to Erd\H{o}s, Ko, and Rado~\cite{EKR61}, concerns the notion of {\em intersecting} set families, where any two sets share at least one element. It states that the largest size of an intersecting family of \mbox{$k$-subsets} of $[n] = \{1,\ldots,n\}$ is $\binom{n-1}{k-1}$, attained by the family of all $k$-subsets that include a fixed element. Another landmark result, proved by Sperner~\cite{Sperner28} in 1928, deals with {\em antichains} --- families of sets with no set containing another --- and asserts that the largest size of an antichain over $[n]$ is $\binom{n}{\lfloor n/2 \rfloor}$, realized by the family of all subsets of $[n]$ of size $\lfloor n/2 \rfloor$. This result has seen various extensions and variations over the years. A notable example is the canonical theorem of Bollob\'as~\cite{Bollobas65}, showing that for all positive integers $a$ and $b$, the maximum size $m$ of a collection $\{(A_i,B_i)\}_{i \in [m]}$ of pairs of sets satisfying $|A_i| = a$ and $|B_i|=b$ for all $i \in [m]$ and $A_i \cap B_j = \emptyset$ if and only if $i=j$ is $\binom{a+b}{a}$. This maximum is achieved by the collection of all pairs $(A,B)$ of subsets of $[a+b]$ with $|A| = a$ and $B = [a+b] \setminus A$.

Significant attention has been devoted in the literature to extremal combinatorics questions concerning permutations. A family of permutations of $[n]$ is called {\em intersecting} if any two of its members agree on at least one element. Analogously to the result of~\cite{EKR61}, Frankl and Deza~\cite{DezaFrankl77} proved in 1977 that the largest size of an intersecting family of permutations of $[n]$ is $(n-1)!$, attained by the collection of all permutations fixing a particular element. This notion naturally extends to that of {\em $t$-intersecting} families, for a positive integer $t \leq n$, where any two permutations are required to agree on at least $t$ elements. It was conjectured in~\cite{DezaFrankl77} that for all integers $t$ and $n$ with $n$ sufficiently greater than $t$, the largest size of a $t$-intersecting family of permutations of $[n]$ is $(n-t)!$. The conjecture was proved more than three decades later independently by Ellis and by Friedgut and Pilpel~\cite{EllisFP11}, and the dependence of $n$ on $t$ has since been considerably sharpened (see, e.g.,~\cite{KellerLMS24}). Another variant of this problem handles {\em $t$-setwise intersecting} families of permutations of $[n]$, in which any two permutations agree on some $t$ elements as a set. Namely, for any two permutations $\pi$ and $\pi'$ in the family, there exists a $t$-subset $A$ of $[n]$ such that $\pi(A) = \pi'(A)$, where $\pi(A)$ stands for the image of $A$ under $\pi$, that is, $\pi(A) = \{\pi(x) \mid x \in A\}$. Ellis conjectured in~\cite{Ellis12} that for all positive integers $t \leq n$, the largest size of a $t$-setwise intersecting family of permutations of $[n]$ is $t! \cdot (n-t)!$, which is realized by the permutations of $[n]$ that map a fixed $t$-subset of $[n]$ to itself. He confirmed the conjecture for the case where $n$ is sufficiently large as a function of $t$. The conjecture was also verified for $t=2$ and for $t=3$ for all admissible values of $n$ by Meagher et al.~\cite{MeagherRaz2021} and Behajaina et al.~\cite{BehajainaMRR21}, respectively.

The results discussed above concern families of permutations that are forced to exhibit a prescribed amount of agreement. A natural complementary objective is to explore the opposite extremal behavior, namely, how large a family of permutations can be if its members are required to disagree in a structured way. One common instance of this requirement is that any two distinct permutations in the family disagree {\em pointwise} on at least some specified number of elements. This corresponds to the concept of {\em permutation codes}, which has been extensively studied in coding theory (see, e.g.,~\cite{CameronPerm}). The present paper investigates a {\em setwise} variant of disagreement among permutations, with applications to the parameterized complexity of graph coloring problems, as described below.

\subsection{Setwise Distinguishable Permutations}

We study the maximum size of a family of permutations of $[n]$ in which each permutation is distinguished from all others by its image on some subset. We call such families {\em setwise distinguishable} and define them formally as follows.

\begin{definition}\label{def:distinguishable}
For a positive integer $n$, a family $\calF$ of permutations of $[n]$ is called {\em setwise distinguishable} if for every permutation $\pi \in \calF$ there exists a set $A_\pi \subseteq [n]$, such that $\pi(A_\pi) \neq \pi'(A_\pi)$ for all $\pi' \in \calF \setminus \{\pi\}$.
\end{definition}

For a positive integer $n$, let $m(n)$ denote the largest possible size of a setwise distinguishable family of permutations of $[n]$.
It is easy to see that $m(n) \leq 2^{2n}$. Indeed, given a setwise distinguishable family $\calF$ of permutations of $[n]$, consider for each permutation $\pi \in \calF$ the pair $(A_\pi, \pi(A_\pi))$, where $A_\pi$ is the subset of $[n]$ distinguishing $\pi$ from all other permutations in $\calF$. Since any two distinct permutations $\pi,\pi' \in \calF$ satisfy $\pi(A_\pi) \neq \pi'(A_\pi)$, it follows that each permutation in $\calF$ is associated with a distinct pair of subsets, implying that $|\calF| \leq 2^{2n}$ (see Lemma~\ref{lemma:upper_m(n)} for a slightly tighter bound). On the other hand, consider a maximal family of permutations of $[n]$ whose images on a fixed subset of size $\lfloor n/2 \rfloor$ are pairwise distinct. Such a family yields that $m(n) \geq \binom{n}{\lfloor n/2 \rfloor} = 2^{(1-o(1)) \cdot n}$, where the $o(1)$ term tends to zero as $n$ tends to infinity. Our main result establishes an exponential improvement over this lower bound, showing that there exists a setwise distinguishable family of permutations of $[n]$ of size $2^{(2-o(1)) \cdot n}$. This yields the following asymptotic estimate for $m(n)$, which is tight up to the $o(1)$ term in the exponent.
\begin{theorem}\label{thm:m(n)}
\[m(n) = 2^{(2-o(1)) \cdot n}.\]
\end{theorem}
\noindent
The lower bound on $m(n)$ is obtained through an explicit elementary construction. For refined upper and lower bounds on $m(n)$, see Theorem~\ref{thm:final_m(n)}.

\subsection{Graph Coloring Kernelization}
Our motivation to study setwise distinguishable families of permutations stems from the area of parameterized complexity (see, e.g.,~\cite{KernelBook19}). This area investigates the computational complexity of decision problems in terms of a quantitative feature of their instances, referred to as a parameter. In the central theme of {\em kernelization}, the objective is to efficiently compress a given instance of a parameterized problem into an equivalent instance whose size is bounded by a function that depends only on the parameter. More concretely, a kernel takes as input a pair $(x,k)$, where $x$ is the actual instance and $k$ is the parameter, and produces in polynomial time an equivalent instance $(x',k')$ satisfying $|x'|+k' \leq h(k)$ for some computable function $h: \N \to \R$. Determining the minimal growth rate of the kernel size $h$ for a given problem is a fundamental task in the field.

For a positive integer $q$, the $\qCol$ problem asks to decide whether a given graph is $q$-colorable, that is, whether there exists a function from its vertex set to the color set $[q]$ assigning distinct colors to the endpoints of every edge. The $\qCol$ problem is well known to be $\NP$-hard when $q \geq 3$, and as such, it has been extensively studied in the framework of parameterized complexity under diverse versions and parameterizations. In particular, Cai~\cite{Cai03} introduced the study of the $\qCol$ problem parameterized by the vertex-deletion distance to a prescribed graph family $\calG$. In this problem, called the {\em $\qCol$ problem on $\calG+k\mathrm{v}$ graphs}, the input consists of a graph $G=(V,E)$ and a set $X \subseteq V$ of $k$ vertices whose removal from $G$ gives a graph that belongs to $\calG$, and the goal is to decide whether $G$ is $q$-colorable, where the parameter is $k$. For example, taking $\calG$ as the graph family $\Empty$ that consists of all edgeless graphs, the parameter expresses the size of a vertex cover in the graph. A series of papers by Jansen and Kratsch~\cite{JansenK13} and by Jansen and Pieterse~\cite{JansenP17,JansenP19color} has shown that for every integer $q \geq 3$, the $\qCol$ problem on $\Empty+k\mathrm{v}$ graphs admits a kernel of size $O(k^{q-1} \cdot \log k)$, whereas for any real $\eps >0$, it admits no kernel of size $O(k^{q-1-\eps})$ unless $\NP \subseteq \coNPpoly$. Since this containment is known to imply the collapse of the polynomial-time hierarchy~\cite{Yap83}, these results essentially settle the kernel complexity of the $\qCol$ problems when parameterized by the vertex cover number. Further extensions are given in~\cite{BerkmanH25,PPRW25}.

The kernel complexity of the $\qCol$ problem on $\calG+k\mathrm{v}$ graphs was systematically studied by Jansen and Kratsch~\cite{JansenK13} for general graph families $\calG$. Interestingly, their results reveal a strong connection between the kernel complexity of the problem and the behavior of graphs in $\calG$ as instances of the so-called $\qListCol$ problem, which asks, given a graph $G$ and a list of available colors from $[q]$ for each vertex, whether $G$ admits a proper coloring respecting these lists. On the positive side, they proved that if a graph family $\calG$ is closed under induced subgraphs and if every $\NO$ instance of $\qListCol$ with a graph from $\calG$ contains a $\NO$ sub-instance on at most $t(q)$ vertices, then the $\qCol$ problem on $\calG+k\mathrm{v}$ graphs admits a kernel with $O(k^{q \cdot t(q)})$ vertices. On the negative side, they proved that if a graph family $\calG$ is closed under disjoint union and if there exists an {\em irreducible} $\NO$ instance of $(q-2)\ListCol$ with a graph in $\calG$ on $t(q) \geq 3$ vertices, that is, a $\NO$ instance that becomes a $\YES$ instance upon removing any vertex, then for any real $\eps >0$, the problem admits no kernel of size $O(k^{t(q)-\eps})$ unless $\NP \subseteq \coNPpoly$. Therefore, for applicable graph families $\calG$, estimating the degree of the polynomial kernel complexity of the $\qCol$ problem on $\calG+k\mathrm{v}$ graphs essentially reduces to analyzing the largest number of vertices in an irreducible $\NO$ instance of $\qListCol$ with a graph from $\calG$.

The authors of~\cite{JansenK13} demonstrated their results on the kernelizability of the $\qCol$ problem on $\calG+k\mathrm{v}$ graphs for several graph families $\calG$. In particular, they considered the family $\USplit$ of all graphs whose connected components are {\em split graphs}, that is, graphs whose vertex set can be partitioned into a clique and an independent set. They showed that every $\NO$ instance of $\qListCol$ with a graph in $\USplit$ contains a $\NO$ sub-instance on at most $2^{2q}+q$ vertices, and concluded that the $\qCol$ problem on $\USplit + k\mathrm{v}$ graphs admits a kernel of size $O(k^{f(q)})$ for a function $f: \N \to \N$ satisfying $f(q) = 2^{(2+o(1)) \cdot q}$. They also showed that there exists an irreducible $\NO$ instance of $\qListCol$ with a split graph on $\binom{q}{\lfloor q/2 \rfloor}+\lfloor q/2 \rfloor$ vertices, yielding that for every integer $q \geq 4$ and any real $\eps>0$, the $\qCol$ problem on $\USplit + k\mathrm{v}$ graphs admits no kernel of size $O(k^{g(q)-\eps})$ for some function $g: \N \to \N$ satisfying $g(q)=2^{(1-o(1)) \cdot q}$, unless $\NP \subseteq \coNPpoly$. This result indicates that the degree of the polynomial kernel complexity of these problems grows exponentially with the number of colors $q$, and yet retains a pronounced exponential multiplicative gap to the upper bound. As an application of our construction of setwise distinguishable families of permutations, we produce an irreducible $\NO$ instance of the $\qListCol$ problem with a split graph on $2^{(2-o(1)) \cdot q}$ vertices. Consequently, we nearly close the gap from~\cite{JansenK13} and obtain an asymptotically tight estimate for the degree of the polynomial kernel complexity of the $\qCol$ problem on $\USplit + k\mathrm{v}$ graphs, as stated below.

\begin{theorem}\label{thm:IntroKernel}
There exists a function $g: \N \to \N$ satisfying $g(q) = 2^{(2-o(1)) \cdot q}$, such that for every sufficiently large integer $q$ and for any real $\eps >0$, the $\qCol$ problem on $\USplit + k\mathrm{v}$ graphs does not admit a kernel of size $O(k^{g(q)-\eps})$ unless $\NP \subseteq \coNPpoly$.
\end{theorem}

We note that the lower bound stated in Theorem~\ref{thm:IntroKernel} trivially extends to the $\qCol$ problems on $\calG + k\mathrm{v}$ graphs for any family $\calG$ containing $\USplit$. This in particular applies to the family of graphs whose connected components are cochordal, that is, edge-complements of graphs in which every cycle of length at least four has a chord. Therefore, as a by-product of Theorem~\ref{thm:IntroKernel}, we strengthen the lower bound of~\cite{JansenK13} on the kernel complexity for this graph family.

\subsection{Outline}
The rest of this short paper is organized as follows. Section~\ref{sec:m(n)} presents our upper and lower bounds on $m(n)$, thereby proving Theorem~\ref{thm:m(n)}. Section~\ref{sec:kernel} discusses the implications for graph coloring kernelization, confirming Theorem~\ref{thm:IntroKernel}. Finally, Section~\ref{sec:conclude} provides concluding remarks and potential avenues for further work.

\section{Setwise Distinguishable Permutations}\label{sec:m(n)}

In this section, we study the largest possible size $m(n)$ of a setwise distinguishable family of permutations of $[n]$ (see Definition~\ref{def:distinguishable}). Recall that for a function $\pi: [n] \to [n]$ and a set $A \subseteq [n]$, $\pi(A)$ denotes the image of $A$ under $\pi$, that is, $\pi(A) = \{\pi(x) \mid x \in A\}$. We also write $\calP(A)$ for the power set of $A$. We begin with the following simple lemma, which provides an alternative characterization of $m(n)$.

\begin{lemma}\label{lemma:m(n)=pairs}
For every positive integer $n$, $m(n)$ is the largest integer $m$ for which there exists a collection $\{(A_i,B_i)\}_{i \in [m]}$ of $m$ pairs of subsets of $[n]$ such that for every $i \in [m]$, there exists a permutation \mbox{$\pi_i:[n] \to [n]$} such that for all $j \in [m]$, it holds that
\[ \pi_i(A_j) = B_j ~~~\mbox{if and only if}~~~ i=j.\]
\end{lemma}

\begin{proof}
Let $m$ and $n$ be positive integers. Suppose first that there exists a collection $\{(A_i,B_i)\}_{i \in [m]}$ of $m$ pairs of subsets of $[n]$ as in the lemma. For each $i \in [m]$, let $\pi_i:[n] \to [n]$ denote the guaranteed permutation, which satisfies $\pi_i(A_i) = B_i$ and $\pi_i(A_j) \neq B_j$ for all $j \in [m] \setminus \{i\}$. It then follows that the family $\{\pi_j \mid j \in [m]\}$ is setwise distinguishable, with each $A_j$ distinguishing $\pi_j$ from all other permutations. Conversely, suppose that there exists a setwise distinguishable family $\{\pi_j \mid j \in [m]\}$ of $m$ permutations of $[n]$. It follows that for each $j \in [m]$, there exists a set $A_j \subseteq [n]$ with $\pi_j(A_j) \neq \pi_i(A_j)$ for all $i \in [m] \setminus \{j\}$. For each $j \in [m]$, define $B_j = \pi_j(A_j)$, and notice that the collection $\{(A_i,B_i)\}_{i \in [m]}$ satisfies for all $i,j \in [m]$ that $\pi_i(A_j) = B_j$ if and only if $i=j$. This establishes the lemma.
\end{proof}

An immediate consequence of Lemma~\ref{lemma:m(n)=pairs} is that $m(n) \leq 2^{2n}$ for all positive integers $n$.
The following lemma slightly strengthens this bound.

\begin{lemma}\label{lemma:upper_m(n)}
For every integer $n \geq 2$, it holds that
\[m(n) \leq \sum_{k=1}^{\lfloor n/2 \rfloor}{{\binom{n}{k}}^2}.\]
Therefore, for every positive integer $n$, it holds that $m(n) < \binom{2n}{n}$.
\end{lemma}

\begin{proof}
For an integer $n \geq 2$, let $m = m(n) \geq 2$, and apply Lemma~\ref{lemma:m(n)=pairs} to obtain a collection $\{(A_i,B_i)\}_{i \in [m]}$ of pairs of subsets of $[n]$ such that for every $i \in [m]$, there exists a permutation $\pi_i:[n] \to [n]$ such that for all $j \in [m]$, it holds that $\pi_i(A_j) = B_j$ if and only if $i=j$. In particular, for every $i \in [m]$, we have $\pi_i(A_i) = B_i$, which implies that $|A_i| = |B_i|$. Since all permutations of $[n]$ share the same image on the empty set, and since $m \geq 2$, it further follows that $|A_i| \geq 1$ for all $i \in [m]$. Additionally, for every permutation $\pi:[n] \to [n]$ and for any two sets $A,B \subseteq [n]$, it holds that $\pi(A) = B$ if and only if their complements satisfy $\pi(\overline{A}) = \overline{B}$. Hence, without loss of generality, we may assume that $|A_i| \leq \lfloor n/2 \rfloor$ for all $i \in [m]$, replacing the pair $(A_i,B_i)$ by $(\overline{A_i},\overline{B_i})$ if needed. Finally, for every $k$-subset $A$ of $[n]$ with $1 \leq k \leq \lfloor n/2 \rfloor$, consider all indices $i \in [m]$ with $A_i = A$. The sets $B_i$ corresponding to these indices are pairwise distinct, because for distinct such indices $i$ and $j$, we have $B_i = \pi_i(A_i) = \pi_i(A_j) \neq B_j$, hence their number does not exceed $\binom{n}{k}$. We thus obtain that $m \leq \sum_{k=1}^{\lfloor n/2 \rfloor}{{\binom{n}{k}}^2}$, as required. In particular, using the identity $\sum_{k=0}^{n}{{\binom{n}{k}}^2} = \binom{2n}{n}$, it follows that $m < \binom{2n}{n}$. By $m(1)=1$, the inequality $m(n) < \binom{2n}{n}$ holds for all positive integers $n$, so the proof is complete.
\end{proof}

We now turn to the core of this section. The following theorem provides an explicit construction of a collection of pairs of subsets of $[n]$ satisfying the property from Lemma~\ref{lemma:m(n)=pairs}. This will be used to establish a nearly tight lower bound on $m(n)$. While the first item of the theorem is not required for this bound, it will play a crucial role in the application discussed in the next section.

\begin{theorem}\label{thm:pairs}
Let $k$ and $n$ be positive integers, such that $2k$ divides $n$.
Then there exists a collection $\{(A_i,B_i)\}_{i \in [m]}$ of \[m = {\binom{2k}{k}}^{n/k-1}\] pairs of subsets of $[n]$, each of size $n/2$, such that
\begin{enumerate}
  \item\label{itm:1} for every permutation $\pi:[n] \to [n]$ there exists an $i \in [m]$ such that $\pi(A_i) = B_i$, and
  \item\label{itm:2} for every $i \in [m]$ there exists a permutation $\pi_i:[n] \to [n]$ such that for all $j \in [m]$, it holds that
  \[\pi_i(A_j) = B_j ~~\mbox{ if and only if}~~ i = j.\]
\end{enumerate}
\end{theorem}

\begin{proof}
For positive integers $k$ and $n$ with $2k$ dividing $n$, denote $t = n/(2k)$, partition $[n]$ into $t$ pairwise disjoint sets $C_1, \ldots, C_t$ of size $2k$ each, and let $C'_1$ be an arbitrary $k$-subset of $C_1$. Define
\[ \calF = \Big \{ (A,B) \in \calP([n]) \times \calP([n]) ~\Big |~ |A \cap C_i| = |B \cap C_i| = k \mbox{ for all }i \in [t] \mbox{ and } A \cap C_1 =C'_1 \Big \}.\]
In words, $\calF$ is the collection of all pairs $(A,B)$ of subsets of $[n]$, such that each of $A$ and $B$ includes precisely $k$ elements from each part $C_i$, and the elements of $A$ in $C_1$ are fixed to those of $C'_1$. Note that for every pair $(A,B) \in \calF$, we have $|A|=|B|=t \cdot k = n/2$. It further holds that \[|\calF| = {\binom{2k}{k}}^{t-1} \cdot {\binom{2k}{k}}^{t} = {\binom{2k}{k}}^{2t-1} = {\binom{2k}{k}}^{n/k-1}.\]
We show that the family $\calF$ satisfies the properties asserted in the theorem.

For Item~\ref{itm:1}, consider any permutation $\pi: [n] \to [n]$, and define a bipartite multigraph $G$ with $t$ vertices $x_1, \ldots, x_t$ on one side and $t$ vertices $y_1, \ldots, y_t$ on the other, as follows. For each $r \in [n]$, add to $G$ an edge, marked by $r$, that connects the vertices $x_i$ and $y_j$ for the indices $i$ and $j$ satisfying $r \in C_i$ and $\pi(r) \in C_j$. Note that each vertex $x_i$ is incident in $G$ with exactly $2k$ edges, marked by the elements of $C_i$, connecting it to the vertices $y_j$ associated with the parts $C_j$ to which the elements of $C_i$ are mapped by $\pi$. Since $\pi$ is a permutation of $[n]$, the multigraph $G$ is $2k$-regular, hence K\H{o}nig's line-coloring theorem implies that it admits a proper edge-coloring with $2k$ colors, and thus a partition of its edge set into $2k$ perfect matchings. The $k$ edges marked by the elements of $C'_1$ belong to $k$ distinct matchings, because they are all incident with the vertex $x_1$. Let $A \subseteq [n]$ denote the set of all marks of the edges in these $k$ matchings, and let $B = \pi(A)$. By construction, we have $|A \cap C_i| = |B \cap C_i| = k$ for all $i \in [t]$ and $A \cap C_1 = C'_1$, implying that $(A,B) \in \calF$. It follows that the permutation $\pi$ admits a pair $(A,B) \in \calF$ such that $\pi(A) = B$, as required.

For Item~\ref{itm:2}, consider a pair $(A,B) \in \calF$, and define a function $\pi : [n] \to [n]$ as follows.
For each $i \in [t]$, define $\pi$ on the elements of $C_i$ so that
\[\pi(A \cap C_i) = B \cap C_i ~~\mbox{and}~~ \pi(C_i \setminus A) = C_{i+1} \setminus B,\]
where the set $C_{t+1}$ is identified with $C_1$. Note that this is possible, because $|A \cap C_i| = |B \cap C_i| = k$ and $|C_i \setminus A| = |C_{i+1} \setminus B| = k$ for all $i \in [t]$. The function $\pi$ is a permutation of $[n]$, because the $2t$ sets $B \cap C_i$ and $C_{i+1} \setminus B$ with $i \in [t]$ form a partition of $[n]$. It further holds that $\pi(A) = B$, because $\pi(A \cap C_i) = B \cap C_i$ for all $i \in [t]$.

It remains to show that for every pair $(A',B') \in \calF$ with $(A',B') \neq (A,B)$, the permutation $\pi$ satisfies $\pi(A') \neq B'$. To this end, fix such a pair $(A',B')$. If $A=A'$, then we have $B \neq B'$, hence $\pi(A') = \pi(A) = B \neq B'$, as desired. So suppose that $A \neq A'$, and recall that we have $A \cap C_1 = A' \cap C_1 = C'_1$. This implies that there exists some $j \in [t-1]$ such that
\[A \cap C_j = A' \cap C_j ~~\mbox{and}~~ A \cap C_{j+1} \neq A' \cap C_{j+1}.\]
Using $|A \cap C_{j+1}| = |A' \cap C_{j+1}| = k$, it follows that there exists an element $r \in (A \cap C_{j+1}) \setminus A'$.
By the definition of $\pi$, the preimage of $C_{j+1}$ under $\pi$ consists of the $k$ elements of $A \cap C_{j+1}$, which are mapped to those of $B \cap C_{j+1}$, and the $k$ elements of $C_j \setminus A$, which are mapped to those of $C_{j+1} \setminus B$.
Therefore, the fact that $r \in (A \cap C_{j+1}) \setminus A'$ implies that $\pi(r) \in (B \cap C_{j+1}) \setminus \pi(A')$.
Further, since $A \cap C_j = A' \cap C_j$, it follows that $C_j \setminus A = C_j \setminus A'$, which by $\pi(C_j \setminus A) = C_{j+1} \setminus B$ implies that the set $\pi(A')$ includes no element of $C_{j+1} \setminus B$, hence $(\pi(A') \cap C_{j+1}) \subseteq (B \cap C_{j+1})$. Consequently, we obtain that $(\pi(A') \cap C_{j+1}) \subseteq (B \cap C_{j+1}) \setminus \{\pi(r)\}$, implying that $|\pi(A') \cap C_{j+1}| \leq k-1$. However, by $(A',B') \in \calF$, we have $|B' \cap C_{j+1}| = k$, yielding that $\pi(A') \neq B'$, as claimed.
\end{proof}

Equipped with Theorem~\ref{thm:pairs}, we obtain the following lower bound on $m(n)$.

\begin{corollary}\label{cor:lower_m(n)}
For all positive integers $k$ and $n$ with $n \geq 2k$, it holds that
\[ m(n) \geq {\binom{2k}{k}}^{2 \cdot \lfloor n/(2k) \rfloor -1}.\]
\end{corollary}

\begin{proof}
For integers $k$ and $n$ with $n \geq 2k$, let $n'$ denote the largest multiple of $2k$ not exceeding $n$. We clearly have $m(n) \geq m(n')$, as any setwise distinguishable family of permutations of $[n']$ can be transformed into a setwise distinguishable family of permutations of $[n]$ with the same size, by fixing the elements of $[n] \setminus [n']$ under all permutations. Combining Theorem~\ref{thm:pairs} with Lemma~\ref{lemma:m(n)=pairs}, we obtain that
\[ m(n) \geq m(n') \geq {\binom{2k}{k}}^{n'/k-1} = {\binom{2k}{k}}^{2 \cdot \lfloor n/(2k) \rfloor-1},\]
so we are done.
\end{proof}

The following statement summarizes our upper and lower bounds on $m(n)$ and confirms Theorem~\ref{thm:m(n)}.
Here and throughout, all logarithms are in base $2$.

\begin{theorem}\label{thm:final_m(n)}
There exists a constant $c > 0$ such that for all positive integers $n$, it holds that
\[ 2^{2n-c \cdot (n \log n)^{1/2}} \leq m(n) \leq 2^{2n-(\log n)/2}.\]
\end{theorem}

\begin{proof}
For the upper bound on $m(n)$, apply Lemma~\ref{lemma:upper_m(n)} to obtain that $m(n) < \binom{2n}{n} \leq \frac{2^{2n}}{\sqrt{n}}$.
For the lower bound, let $n$ be a positive integer, and let $k = \lfloor (n \log n)^{1/2} \rfloor$. We may assume that $n$ is sufficiently large, and thus $k \geq 1$ and $n \geq 2k$, as smaller values of $n$ can be handled by an appropriate choice for the constant $c$ using $m(n) \geq 1$. By Corollary~\ref{cor:lower_m(n)}, we obtain that
\begin{eqnarray*}
m(n) & \geq & {\binom{2k}{k}}^{2 \cdot \lfloor n/(2k) \rfloor -1} \geq {\binom{2k}{k}}^{2 \cdot (n/(2k) -1) -1}
\geq \bigg ( \frac{2^{2k}}{2 \sqrt{k}} \bigg )^{n/k-3} \\ & = & 2^{(2k-(\log k)/2-1) \cdot (n/k-3)}
\geq 2^{2n-6k-(n\log k)/(2k)-n/k} = 2^{2n - \Theta((n \log n)^{1/2})}.
\end{eqnarray*}
This completes the proof.
\end{proof}

We conclude this section by highlighting a limitation of a natural approach to constructing setwise distinguishable families of permutations of $[n]$. Such a family can be obtained from a collection $\{(A_i,B_i)\}_{i \in [m]}$ of $m$ pairs of subsets of $[n]$ with the property that for all $i,j \in [m]$, it holds that $|A_i \cap A_j| = |B_i \cap B_j|$ if and only if $i=j$. Indeed, {\em any} permutation $\pi_i$ of $[n]$ that maps $A_i$ to $B_i$ necessarily satisfies $\pi_i(A_j) \neq B_j$ for every $j \in [m] \setminus \{i\}$, as otherwise it would follow that
\[|A_i \cap A_j| = |\pi_i(A_i \cap A_j)| = |\pi_i(A_i) \cap \pi_i(A_j)| = |B_i \cap B_j|,\]
contradicting the prescribed intersection property. This yields, by Lemma~\ref{lemma:m(n)=pairs}, a setwise distinguishable family of size $m$. For example, by fixing all sets $A_i$ to the same subset of $[n]$ of size $\lfloor n/2 \rfloor$ and varying $B_i$ over all subsets of the same size, one obtains a collection of $\binom{n}{\lfloor n/2 \rfloor}$ pairs with the desired property. The following result shows, however, that no larger such collection exists. Its proof is inspired by the classic probabilistic argument underlying Sperner's theorem (see, e.g.,~\cite[Chapter~12]{AlonS16}).

\begin{proposition}
For positive integers $m$ and $n$, let $\{(A_i,B_i)\}_{i \in [m]}$ be a collection of $m$ pairs of subsets of $[n]$ such that for all $i,j \in [m]$, it holds that
\[ |A_i \cap A_j| = |B_i \cap B_j| ~~~\mbox{if and only if}~~~i=j.\]
Then \[m \leq \binom{n}{\lfloor n/2 \rfloor}.\]
\end{proposition}

\begin{proof}
Let $\{(A_i,B_i)\}_{i \in [m]}$ be a collection of $m$ pairs of subsets of $[n]$ as in the statement. It follows that $|A_i|=|B_i|$ for all $i \in [m]$. Let $\pi$ be a permutation of $[n]$ chosen uniformly at random. For each $i \in [m]$, let $E_i$ denote the event that $\pi(A_i) = B_i$. Since $\pi(A_i)$ is uniformly distributed over all subsets of $[n]$ of size $|A_i|$, we have $\Prob{}{E_i} = 1/{\binom{n}{|A_i|}}$. We claim that the events $E_i$ with $i \in [m]$ are pairwise disjoint. Indeed, suppose that $E_i$ and $E_j$ both occur, so that $\pi(A_i) = B_i$ and $\pi(A_j)=B_j$. Since $\pi$ is a permutation of $[n]$, this implies that $\pi(A_i \cap A_j) = \pi(A_i) \cap \pi(A_j) = B_i \cap B_j$, which in turn implies that $|A_i \cap A_j| = |B_i \cap B_j|$, and therefore $i=j$. We conclude that
\begin{eqnarray}\label{eq:E_i}
1 \geq \Prob{}{\bigvee_{i =1}^{m}{E_i}} = \sum_{i=1}^{m}{\Prob{}{E_i}} = \sum_{i=1}^{m}{\frac{1}{\binom{n}{|A_i|}}} \geq m \cdot \frac{1}{\binom{n}{\lfloor n/2 \rfloor}},
\end{eqnarray}
where the first equality holds because the events $E_i$ are pairwise disjoint, and the last inequality holds because $\binom{n}{k} \leq \binom{n}{\lfloor n/2 \rfloor}$ for all $0 \leq k \leq n$. Rearranging~\eqref{eq:E_i} yields that $m \leq \binom{n}{\lfloor n/2 \rfloor}$, completing the proof.
\end{proof}

\section{Graph Coloring Kernelization}\label{sec:kernel}

In this section, we employ our construction of setwise distinguishable families of permutations to obtain improved lower bounds on the kernel complexity of coloring problems parameterized by the vertex-deletion distance to split graphs, thereby proving Theorem~\ref{thm:IntroKernel}. We first collect a few related definitions and concepts.

\subsection{Preliminaries}

\paragraph*{Kernelization.}
A {\em parameterized problem} is a set $Q \subseteq \Sigma^* \times \N$ for some finite alphabet $\Sigma$. A {\em kernelization} (or simply a kernel) for $Q$ is an algorithm that, given an instance $(x,k) \in \Sigma^* \times \N$, returns in time polynomial in $|x|+k$ an instance $(x',k') \in \Sigma^* \times \N$, such that $(x,k) \in Q$ if and only if $(x',k') \in Q$, and such that $|x'|+k' \leq h(k)$ for some computable function $h:\N \to \R$. The function $h$ is called the {\em size} of the kernel.
For an in-depth introduction to the area of kernelization, the reader is referred to~\cite{KernelBook19}.

\paragraph*{Graph Coloring.}
For a positive integer $q$, a {\em $q$-coloring} of a graph $G=(V,E)$ is a function \mbox{$c: V \to [q]$}, which assigns a color $c(v)$ to each vertex $v \in V$. The coloring $c$ of $G$ is called {\em proper} if it satisfies $c(u) \neq c(v)$ for every edge $\{u,v\} \in E$. For a list function $L: V \to \calP([q])$, a {\em proper list coloring} of the pair $(G,L)$ is a proper $q$-coloring $c: V \to [q]$ of $G$ satisfying $c(v) \in L(v)$ for all $v \in V$. For a fixed integer $q$, the $\qCol$ problem asks whether a given graph $G$ is $q$-colorable, and the $\qListCol$ problem asks whether a given pair $(G,L)$, consisting of a graph $G=(V,E)$ and a list function $L:V \to \calP([q])$, admits a proper list coloring. A {\em sub-instance} of $(G,L)$ is an instance obtained from $(G,L)$ by removing a subset of the vertex set of $G$ and restricting $L$ to the remaining vertices. A $\NO$ instance of $\qListCol$ is called {\em irreducible} if removing any vertex results in a $\YES$ sub-instance.

For a positive integer $q$ and for a graph family $\calG$, the {\em $\qCol$ problem on $\calG + k\mathrm{v}$ graphs} is the parameterized problem with parameter $k$ that, given a graph $G=(V,E)$ and a set $X \subseteq V$ of size $k$ whose removal from $G$ gives a graph in $\calG$, asks to decide whether $G$ is $q$-colorable. The following theorem, proved by Jansen and Kratsch~\cite{JansenK13} using machinery developed by Dell and van Melkebeek~\cite{DellM14}, shows that conditional lower bounds on the kernel complexity of this problem can be derived from irreducible $\NO$ instances of the $(q-2)\ListCol$ problem with graphs from $\calG$.

\begin{theorem}[{\cite[Theorem~4]{JansenK13}}]\label{thm:JK_lower}
Let $\calG$ be a graph family closed under disjoint union, and let $q \geq 3$ be a positive integer such that there exists an irreducible $\NO$ instance of $(q-2)\ListCol$ with a graph from $\calG$ on $t \geq 3$ vertices. Then, for any real $\eps >0$, the $\qCol$ problem on $\calG + k\mathrm{v}$ graphs admits no kernel of size $O(k^{t-\eps})$, and thus no kernel with $O(k^{(t-\eps)/2})$ vertices, unless $\NP \subseteq \coNPpoly$.
\end{theorem}

A {\em split graph} is a graph whose vertex set can be partitioned into a clique and an independent set. Let $\Split$ denote the family of all split graphs, and let $\USplit$ denote the family of all graphs whose connected components are members of $\Split$. The family $\USplit$ is clearly closed under disjoint union.

\subsection{Irreducible NO Instances of List Coloring on Split Graphs}

The following theorem produces an irreducible $\NO$ instance of the $\qListCol$ problem on split graphs, surpassing the number of vertices achieved in~\cite[Lemma~6]{JansenK13}.

\begin{theorem}\label{thm:NO_split}
For all positive integers $k$ and $q$ with $q \geq 2k$, there exists an irreducible $\NO$ instance $(G,L)$ of $\qListCol$, where $G$ is a split graph on at least \[{\binom{2k}{k}}^{2 \cdot \lfloor q/(2k) \rfloor-1}\]
vertices. Therefore, there exists a constant $c>0$ such that for all positive integers $q$, there exists such an instance on at least $2^{2q-c \cdot (q \log q)^{1/2}}$ vertices.
\end{theorem}

\begin{proof}
For the first part of the theorem, let $k$ and $q$ be positive integers satisfying $q \geq 2k$. We may assume without loss of generality that $2k$ divides $q$, since otherwise, one can apply the result with $q$ replaced by the largest multiple of $2k$ not exceeding it, yielding a valid instance of $\qListCol$ that satisfies the stated bound on the number of vertices. Under this assumption, let $m = {\binom{2k}{k}}^{q/k-1}$, and apply Theorem~\ref{thm:pairs} to obtain a collection $\{(A_i,B_i)\}_{i \in [m]}$ of $m$ pairs of subsets of $[q]$, each of size $q/2$, such that
\begin{enumerate}
  \item\label{itm:1_NO} for every permutation $\pi:[q] \to [q]$ there exists an $i \in [m]$ such that $\pi(A_i) = B_i$, and
  \item\label{itm:2_NO} for every $i \in [m]$ there exists a permutation $\pi_i:[q] \to [q]$ such that for all $j \in [m]$, it holds that $\pi_i(A_j) = B_j$ if and only if $i = j$.
\end{enumerate}

Let $C = \cup_{i \in [m]}{A_i} \subseteq [q]$. We define a graph $G=(V,E)$ on the vertex set $V = C \cup \{v_1, \ldots, v_m\}$, where the vertices of $C$ are pairwise adjacent, and for each $i \in [m]$, the vertex $v_i$ is adjacent to the vertices of $A_i \subseteq C$. Clearly, $G$ is a split graph, as $C$ forms a clique, and $\{v_1, \ldots, v_m\}$ forms an independent set. Furthermore, the number of vertices in $G$ satisfies $|V| \geq m  = {\binom{2k}{k}}^{q/k-1}$, which matches the desired bound assuming that $2k$ divides $q$. We define a list function $L:V \to \calP([q])$ as follows. For each $r \in C$, let $L(r)=[q]$, and for each $i \in [m]$, let $L(v_i) = B_i \subseteq [q]$. It remains to show that $(G,L)$ is an irreducible $\NO$ instance of the $\qListCol$ problem.

We first show that $(G,L)$ is a $\NO$ instance of $\qListCol$. Let $c:V \to [q]$ be a coloring of $G$ satisfying $c(v) \in L(v)$ for all $v \in V$. We show that $c$ is not proper. If $c$ assigns the same color to two distinct vertices in the clique $C$, then it is clearly not proper. Otherwise, there exists a permutation $\pi:[q] \to [q]$ that agrees with $c$ on the vertices of $C$, meaning that $\pi(r) = c(r)$ for all $r \in C$. By Item~\ref{itm:1_NO} above, there exists an $i \in [m]$ such that $\pi(A_i) = B_i$, and thus $c(A_i)=B_i$. However, the vertex $v_i$ is adjacent in $G$ to all vertices of $A_i \subseteq C$, and its color $c(v_i)$ lies in $L(v_i) = B_i$. Therefore, there exists a vertex $r \in A_i$ with $c(r)=c(v_i)$, showing that $c$ is not proper, as desired.

We next show that removing any vertex from $G$ yields a $\YES$ sub-instance of $(G,L)$ for the $\qListCol$ problem.
Consider first an instance obtained from $(G,L)$ by removing a vertex $v_i$ for some $i \in [m]$. By Item~\ref{itm:2_NO} above, there exists a permutation $\pi_i:[q] \to [q]$ such that for all \mbox{$j \in [m]$}, it holds that $\pi_i(A_j) = B_j$ if and only if $i = j$. We define a coloring $c$ of the vertices of $V \setminus \{v_i\}$ as follows. For each vertex $r \in C$, let $c(r) = \pi_i(r)$. Since $\pi_i$ is a permutation of $[q]$, the endpoints of every edge in the clique $C$ are assigned distinct colors under $c$. Next, for each $j \in [m] \setminus \{i\}$, it holds that $\pi_i(A_j) \neq B_j$, and thus $c(A_j) \neq B_j$, which using $|c(A_j)| = |A_j| = |B_j|$ implies that $B_j \setminus c(A_j) \neq \emptyset$. This allows us to assign to the vertex $v_j$ a color $c(v_j)$ from its list $L(v_j)=B_j$ that is not assigned by $c$ to any vertex in its neighborhood $A_j$. The obtained coloring $c$ is proper and respects the list function $L$, as required.

Finally, consider an instance obtained from $(G,L)$ by removing some vertex $r \in C$. By the definition of $C$, there exists an $i \in [m]$ such that $r \in A_i$. By the preceding paragraph, the instance obtained from $(G,L)$ by removing the vertex $v_i$ admits a proper list coloring $c$. Since $(G,L)$ is a $\NO$ instance, this coloring must satisfy $c(A_i) = B_i$, as otherwise, using $|c(A_i)|= |A_i| = |B_i|$, it could be extended to a proper list coloring of the entire instance $(G,L)$. For the instance $(G,L)$ with the vertex $r$ removed, we use the coloring $c$ on the vertices of $V \setminus \{r,v_i\}$ and assign the color $c(r)$ to the vertex $v_i$. Since $r \in A_i$ and $c(A_i) = B_i$, the color $c(r)$ lies in $L(v_i) = B_i$ and is absent from the neighborhood of $v_i$ when $r$ is removed, yielding a proper list coloring for the resulting sub-instance. This completes the proof of the first part of the theorem.

For the second part of the theorem, it may be assumed that $q$ is sufficiently large, as for smaller values of $q$ and for an appropriate choice of the constant $c>0$, one may take a graph with a single vertex assigned an empty set of colors. Under this assumption, the result follows by applying the first part of the theorem with $k = \lfloor (q \log q)^{1/2} \rfloor$. We omit the calculation, which is identical to the one in the proof of Theorem~\ref{thm:final_m(n)}.
\end{proof}

We complement Theorem~\ref{thm:NO_split} with the following simple upper bound on the size of any irreducible $\NO$ instance of the $\qListCol$ problem on split graphs. It slightly improves the $2^{2q}+q$ bound obtained in~\cite[Lemma~2]{JansenK13}.

\begin{proposition}\label{prop:irreducible_upper}
For every positive integer $q$, every $\NO$ instance of the $\qListCol$ problem with a graph in $\USplit$ has a $\NO$ sub-instance on at most $2^q \cdot \binom{q}{\lfloor q/2 \rfloor} + q$ vertices.
\end{proposition}

\begin{proof}
For a positive integer $q$, let $(G,L)$ be a $\NO$ instance of the $\qListCol$ problem, where $G=(V,E)$ is a graph in the family $\USplit$, and $L:V \to \calP([q])$ is a list function. We may assume that $G$ belongs to $\Split$, since otherwise, some connected component of $G$ already induces a $\NO$ sub-instance of $(G,L)$. It follows that there exists a partition $V = C \cup I$ of the vertex set of $G$ into a clique $C$ and an independent set $I$. If $|C| \geq q+1$, then any $q+1$ vertices of $C$ induce a $\NO$ sub-instance of $(G,L)$, so we are done. Otherwise, it holds that $|C| \leq q$, and we associate with each vertex $v \in I$ its neighborhood $N(v) \subseteq C$. As long as there exist two distinct vertices $v_1,v_2 \in I$ with $N(v_1) = N(v_2)$ and $L(v_1) \subseteq L(v_2)$, we remove the vertex $v_2$ from the set $I$ and from the graph $G$. Such a removal results in a $\NO$ sub-instance, because any color assignment to the vertex $v_1$ also suits $v_2$. At the end of this process, for each set $A \subseteq C$, the collection of lists $\{L(v) \mid v \in I,~N(v) = A\}$ forms an antichain, that is, it includes no two sets one containing the other, so by Sperner's theorem~\cite{Sperner28}, its size does not exceed $\binom{q}{\lfloor q/2 \rfloor}$. Since the number of subsets of $C$ is bounded by $2^q$, we obtain that the total number of vertices in the produced $\NO$ sub-instance of $(G,L)$ is at most $2^q \cdot \binom{q}{\lfloor q/2 \rfloor} + |C| \leq 2^q \cdot \binom{q}{\lfloor q/2 \rfloor} + q$, so we are done.
\end{proof}

\subsection{Proof of Theorem~\ref{thm:IntroKernel}}

We are now ready to prove Theorem~\ref{thm:IntroKernel}, which asserts that there exists a function \mbox{$g: \N \to \N$} with $g(q) = 2^{(2-o(1)) \cdot q}$, such that for every sufficiently large integer $q$ and for any real $\eps >0$, the $\qCol$ problem on $\USplit + k\mathrm{v}$ graphs does not admit a kernel of size $O(k^{g(q)-\eps})$ unless $\NP \subseteq \coNPpoly$.

\begin{proof}[of Theorem~\ref{thm:IntroKernel}]
By Theorem~\ref{thm:NO_split}, there exists a function $g: \N \to \N$ with $g(q) = 2^{(2-o(1)) \cdot q}$, such that for every integer $q \geq 3$, there exists an irreducible $\NO$ instance of $(q-2)\ListCol$ whose graph lies in the family $\USplit$ and has at least $g(q)$ vertices. Since $\USplit$ is closed under disjoint union and $g(q) \geq 3$ for every sufficiently large integer $q$, we can apply Theorem~\ref{thm:JK_lower} and obtain that for every such $q$ and any real $\eps >0$, the $\qCol$ problem on $\USplit + k\mathrm{v}$ graphs does not admit a kernel of size $O(k^{g(q)-\eps})$ unless $\NP \subseteq \coNPpoly$. The proof is now complete.
\end{proof}

\section{Concluding Remarks}\label{sec:conclude}

This paper investigates the maximum size $m(n)$ of a setwise distinguishable family of permutations of $[n]$. Theorem~\ref{thm:final_m(n)} shows that for some constant $c > 0$ and for all positive integers $n$, it holds that
\[ 2^{2n-c \cdot (n \log n)^{1/2}} \leq m(n) \leq 2^{2n-(\log n)/2}.\]
These upper and lower bounds coincide up to the lower-order terms in the exponent. It would be intriguing to close the remaining gap.

Despite the purely combinatorial nature of the quantity $m(n)$, our motivation for exploring it arose from parameterized complexity, specifically, in the study of the kernel complexity of graph coloring problems parameterized by the vertex-deletion distance to split graphs, addressed by Jansen and Kratsch~\cite{JansenK13}. Our construction of setwise distinguishable families of permutations allowed us to clarify the exponential growth rate of the degree of the polynomial kernel size of these problems as a function of the number of colors, assuming a standard complexity assumption. We also strengthened the conditional lower bounds on the kernel complexity of coloring problems parameterized by the vertex-deletion distance to the family of graphs whose connected components are cochordal, considered in~\cite{JansenK13}. For this family, however, the gap to the best known upper bound remains substantial, and narrowing it constitutes a natural direction for further research.

\section*{Acknowledgments}
We would like to thank Shoni Gilboa for fruitful discussions and the anonymous reviewers for useful feedback.

\bibliographystyle{abbrv}
\bibliography{perm-inter}

\end{document}